\documentstyle[epsf]{aipproc}

\def\alt{\mathrel{\raise.2ex\hbox{$<$}\hskip-.8em\lower.9ex\hbox{$\sim$}}}
\def\agt{\mathrel{\raise.2ex\hbox{$>$}\hskip-.8em\lower.9ex\hbox{$\sim$}}}

\def\cstar{\cos\theta^*}
\def\mVV{M_{VV}}
\def\rts{\sqrt s}

\def\h{H}           
\def\mh{m_{\h}}
\def\mt{m_t}

\def\fbi{~{\rm fb}^{-1}}

\def\gev{~{\rm GeV}}
\def\tev{~{\rm TeV}}
\def\anti{\overline}

\def\eps{\epsilon}
\def\stop{\widetilde t}
\def\mstop{m_{\stop}}
\def\hl{h^0}

\def\sighbar{\overline \sigma_{\h}}

\def\mbb{m_{b\anti b}}

\def\shat{{\hat s}}
\def\rtshat{\sqrt{\shat}}

\def\sighbar{\overline \sigma_{\h}}

\def\anti{\overline}

\def\rts{\sqrt s}

\def\eps{\epsilon}
\def\anti{\overline}
\def\wp{W^+}

\def\h{h}
\def\mh{m_{\h}}

\def\hsm{h_{SM}}
\def\mhsm{m_{\hsm}}

\def\tanb{\tan\beta}
\def\hl{h^0}

\def\ha{A^0}
\def\mha{m_{\ha}}

\def\hh{H^0}

\def\fbi{~{\rm fb}^{-1}}

\def\gev{~{\rm GeV}}
\def\tev{~{\rm TeV}}
\def\stop{\widetilde t}
\def\mstop{m_{\stop}}
\def\mt{m_t}

 \catcode`@=11
\def\@cite#1#2{\unskip\nobreak\relax
    \def\@tempa{$\m@th^{\hbox{\the\scriptfont0 #1}}$}%
    \futurelet\@tempc\@citexx}
\def\@citexx{\ifx.\@tempc\let\@tempd=\@citepunct\else
    \ifx,\@tempc\let\@tempd=\@citepunct\else
    \let\@tempd=\@tempa\fi\fi\@tempd}
\def\@citepunct{\@tempc\edef\@sf{\spacefactor=\the\spacefactor\relax}\@tempa
    \@sf\@gobble}
 
\def\citenum#1{{\def\@cite##1##2{##1}\cite{#1}}}
\def\citea#1{\@cite{#1}{}}
\catcode`@=12 

\begin{document}
\thispagestyle{empty}

\font\fortssbx=cmssbx10 scaled \magstep2
\hbox to \hsize{
\hbox{\fortssbx University of Wisconsin - Madison}
\hfill$\vbox{\hbox{\bf MADPH-97-989}
                \hbox{March 1997}}$ }

\title{The Physics Capabilities\\[2mm] of $\mu^+\mu^-$ Colliders\footnote{Talk presented by V.~Barger at the symposium \,{\it New Ideas for Particle Accelerators}, Institute for Theoretical Physics, Santa Barbara, California, October 1996.}}
\author{{\it The Muon Quartet Collaboration}\\
V.~Barger$^a$, M.S.~Berger$^b$, J.F.~Gunion$^c$, T.~Han$^c$}

\address{
$^a$Physics Department, University of Wisconsin, Madison, WI 53706,
USA\\
$^b$Physics Department, Indiana University, Bloomington, IN 47405,
USA\\
$^c$Physics Department, University of California,  Davis, CA 95616,  
USA}

\maketitle

\begin{abstract}
We summarize the potential of muon colliders to probe fundamental physics. $W^+W^-$, $\bar tt$, and $Zh$ threshold measurements could determine masses to precisions $\Delta M_W = 6$~MeV, $\Delta m_t = 70$~MeV, and $\Delta m_h = 45$~MeV, to test electroweak radiative corrections. With $s$-channel Higgs production, unique to a muon collider, the Higgs mass could be pinpointed ($\Delta m_h < 1$~MeV) and its width measured. The other Higgs bosons of supersymmetry can be produced and studied by three methods. If instead the $WW$ sector turns out to be strongly interacting, a 4~TeV muon collider is ideally suited to its study.
\end{abstract}
\thispagestyle{empty}

\section{Introduction}

In this report we address the exciting physics that could be accomplished at muon colliders in the context of the central physics issue of our time: how is the electroweak symmetry broken, weakly or strongly? Higgs bosons and a low energy supersymmetry (SUSY) are the particles of interest in the weakly broken scenario and new resonances at the TeV scale of a new strong interaction dynamics are the alternatives.

Muon colliders would have decided advantages over other machines in providing
(i)~sharp beam energy for precision measurements of masses, widths and couplings of the Higgs, $W$, $t$ and supersymmetry  particles, and (ii)~high energy / high luminosity for production of high mass particles and studies of a strongly interacting electroweak sector (SEWS).

In order to be able to do interesting physics at a muon collider, the minimum luminosity requirement is
\begin{equation}
L > \rm {1000\ events/year\over\sigma_{QED}} \,, \label{eq:minlum}
\end{equation}
where $\sigma_{\rm QED}$ is the cross section for the process $\mu^+\mu^-\to\gamma\to e^+e^-$,
\begin{equation}
\sigma_{\rm QED} \approx {100{\rm\ fb}\over s(\rm TeV)^2} \,.
\end{equation}
The prototype designs under consideration well exceed the figure of merit in (\ref{eq:minlum}).

\begin{itemize}
\item \underline{First Muon Collider (FMC)}\\[3mm]
\begin{eqnarray*}\noalign{\vskip-4ex}
&& (250\rm\ GeV)\times (250\ GeV) \qquad
{\cal L} = 2\times10^{33}  cm^{-2} \, s^{-1} \quad (20\ fb^{-1}/year)\,,\\
&& N_{\rm QED} \approx 8000\rm\ events/year\,.
\end{eqnarray*}

\item \underline{Next Muon Collider (NMC)}\\[3mm]
\begin{eqnarray*}\noalign{\vskip-4ex}
&& (2\rm\ TeV)\times (2\ TeV) \qquad
{\cal L} = 10^{35} \, cm^{-2} \, s^{-1} \quad (1000\ fb^{-1}/year) \,,\\
&& N_{\rm QED} \approx 6000\rm\ events/year\,.
\end{eqnarray*}
\end{itemize}

\noindent
Special purpose rings may be added to these designs at modest cost, to optimize luminosities at specific energies for study of $s$-channel resonances and thresholds. 

Muon colliders offer several unique and highly advantageous features. First, $s$-channel Higgs boson production occurs at interesting rates. The Higgs coupling is proportional to the mass, so this process is highly suppressed at $e^+e^-$ and $pp$ colliders. Second, a fine energy resolution is an intrinsic property of muon colliders. A beam energy resolution $R=0.04$ to 0.08\% is natural and a resolution down to $R=0.01\%$ is realizable\cite{private}. By comparison, $R>1\%$ is expected at an $e^+e^-$ machine. The root mean square spread in center-of-mass energy $\sigma_{\sqrt s}$ is given in terms of $R$ by
\begin{equation}
\sigma_{\sqrt s} = (7{\rm\ MeV}) \left(R\over 0.01\%\right) \left(\sqrt s\over 100\rm\ GeV\right) \,.
\end{equation}
The monochromicity of the c.m.\ energy is {\em vital} for $s$-channel Higgs studies and {\em valuable}  for threshold measurements. Furthermore, for mass measurements a c.m.\ calibration can be obtained of MeV accuracy, and the necessary energy calibration $\delta\sqrt s \sim 10^{-6}$ may be achieved with spin rotation measurements of polarized muons in the ring. A final, but no less significant aspect of muon colliders, is that their c.m.\ energy reach is extendable into the $\sqrt s > 1$~TeV range where new physics is expected (either SUSY or SEWS).

\section{Threshold physics at the FMC}

Precision measurements of $M_W$ and $m_t$ can provide important constraints on the Higgs mass in the SM, or other new physics beyond the SM, through the relation
\begin{equation}
M_W = M_Z \left[ 1 - {\pi\alpha\over \sqrt 2 \, G_\mu \, M_W^2 (1-\delta r)} \right] ^{1/2} \,,
\end{equation}
where the loop contributions $\delta r$ depend on $m_t^2$ and $\log m_h$ in the SM and also on sparticle masses in supersymmetry. The optimal relative precision of $M_W$ and $m_t$ measurements is
\begin{equation}
\Delta M_W \approx {1\over 140} \Delta m_t \,,
\end{equation}
for example $\Delta M_W\approx 2$~MeV for $\Delta m_t\approx 200$~MeV.

\subsection{$WW$ and $t\bar t$ thresholds\protect\cite{mwmt}}

There is excellent potential to make very precise $M_W$, $m_t$ and $m_h$ measurements at the FMC because of the sharp beam energy, the suppression of initial state radiation and the absence of significant beamstrahlung. With 100~fb$^{-1}$ luminosity just above the $WW$ threshold ($\sqrt s \approx 161$~GeV) a $M_W$ precision
\begin{equation}
\Delta M_W = 6\rm\ MeV
\end{equation}
is attainable. With 100~fb$^{-1}$ devoted to a 10 point scan ($\sqrt s = 1$~GeV) over the $t\bar t$ threshold, the top quark mass could be measured to an accuracy
\begin{equation}
\Delta m_t = 70\rm\ MeV \,.
\end{equation}
Then, with improved determinations of $\alpha$, $\alpha_s$, and $\sin^2\theta_w$, such $M_W$ and $m_t$ measurements would constrain the SM Higgs to a precision
\begin{equation}
{m_h}_{-0.11m_h}^{+0.13m_h}\,,
\end{equation}
(e.g.\ $100_{-11}^{+13}$~GeV). The limiting factor on $\Delta m_h$ is the uncertainty on $M_W$. Correspondingly tight constraints would be placed on other new physics. The shape of the $t\bar t$ threshold also constrains $\alpha_s(M_Z)$, the top quark width $\Gamma_t$, and possibly also $m_h$ (via $h$-exchange contributions). 

\subsection{$Zh$ threshold\protect\cite{zh}}

In the minimal standard model (MSSM), the parameters of the Higgs sector are 
\begin{equation}
\tan\beta = {v_2\over v_1}\quad{\rm and}\quad m_A
\end{equation} 
at tree level, and also the top-quark and stop masses and stop mixing at the loop level. The light Higgs boson is the only supersymmetric particle with an iron-clad upper bound, which is\cite{carena}
\begin{eqnarray}
m_h < 64\mbox{--105 GeV} &\quad&\rm for\ \tan\beta\simeq\phantom01.8\,,\\
m_h < 98\mbox{--125 GeV} &\quad&\rm for\ \tan\beta\simeq60 \,.
\end{eqnarray}
Thus it is the first target for experimental searches. 

A light Higgs boson will be produced at a lepton collider via the $Z$-bremsstrahlung process
\begin{equation}
\ell^+\ell^- \to Z^* \to Zh\,.
\end{equation}
With high resolution energy measurements the recoil mass and $h$-mass reconstruction give high precision $m_h$. With 50~fb$^{-1}$ luminosity at a $\sqrt s = 500$~GeV $e^+e^-$ collider, the following precisions are anticipated
\begin{eqnarray}
&& \Delta m_h = 180\mbox{ MeV\qquad (SLD type detector)}\,,\\
&& \Delta m_h = \phantom020\mbox{ MeV\qquad (super-JLC detector)} \,.
\end{eqnarray}

A threshold measurement of the $Zh$ production cross section provides a new approach to precisely determine $m_h$. Figure~1 shows the predicted energy dependence of the cross section. The $Zb\bar b$ background is very small, except in the case when $m_h\sim M_Z$. A measurement with 100~fb$^{-1}$ luminosity at $\sqrt s = m_h - M_Z + 0.5$~GeV of the cross section at a muon collider would yield a SM Higgs mass to within
\begin{eqnarray}
\Delta m_h = 45\rm\ MeV 
\end{eqnarray}
for $m_h=100$~GeV.

\begin{figure}[t]
\centering
\hspace{0in}\epsfxsize=4in\epsffile{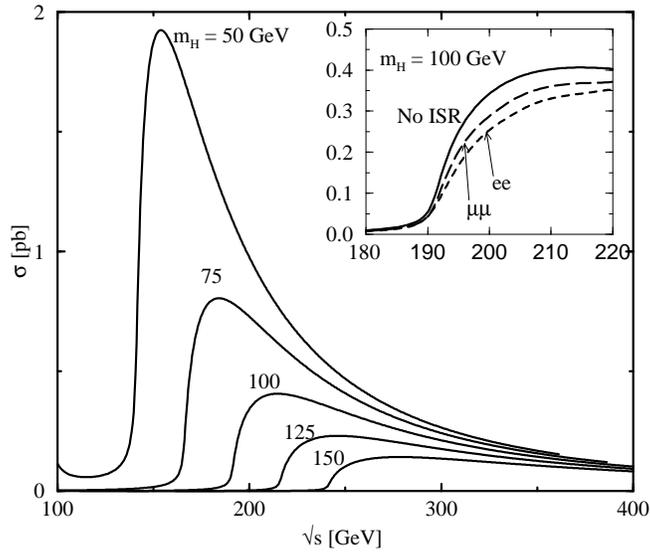}

\caption{The cross section vs.\ $\protect\rts$
for the process
$\ell^+\ell^-\to Z^\star\h\to f\anti f\h$ for a range of Higgs masses. 
The inset figure shows the
detailed structure for $\mh=100$~GeV in the threshold region.
Also shown in the inset figure are the effects
of initial state radiation (ISR) and beam energy smearing assuming
a Gaussian spread $R_e=1\%$ for $e^+e^-$ and
$R_\mu=0.1\%$ for $\mu^+\mu^-$.}
\end{figure}

\section{\lowercase{$s$}-channel Higgs physics\lowercase{\protect\cite{s-channel}}}

A unique capability of a muon collider is the production of a Higgs resonance in the $s$-channel, $\mu^+\mu^-\to h\to b\bar b$. The light quark background can be rejected by $b$-tagging. The resonance cross section is
\begin{equation}
\sigma_h = {4\pi\Gamma(h\to \mu\bar\mu) \; \Gamma(h\to b\bar b)\over
\left(s-m_h^2\right)^2 + m_h^2 \Gamma_h^2} \,.
\end{equation}
One needs to tune the Higgs energy to $\sqrt s = m_h$ by an energy scan in the vicinity of $m_h$. The signal is enhanced with polarized beams if the luminosity decrease with polarization\cite{parsa} is less than $(1+P)^2/(1-P)^2$ which is 10 for $P=0.84$.

The Higgs resonance profile depends on the total Higgs width, $\Gamma_h$, which is highly model dependent. Figure~2 shows expectations for $\Gamma_h$ in the SM and the MSSM. A light ($\alt125$~GeV) SM Higgs has a relatively narrow width, $\Gamma_h \sim{}$few MeV. The width of the light Higgs in the MSSM is larger than that of the SM Higgs and scales up roughly with $(\tan\beta)^2$. Figure~3 shows the light Higgs resonance profile for the SM Higgs and the MSSM Higgs at $\tan\beta = 10$ and 20, for resolutions $R=0.01\%$, 0.06\% and 0.1\%. For a resolution $\sigma\sim\Gamma_h$, the Breit-Wigner line shape can be measured and $\Gamma_h$ determined. To be sensitive to $\Gamma_h$ of a few MeV, a resolution $R=0.01\%$ is needed.

\begin{figure}[t]
\centering
\hspace{0in}\epsfxsize=3.8in\epsffile{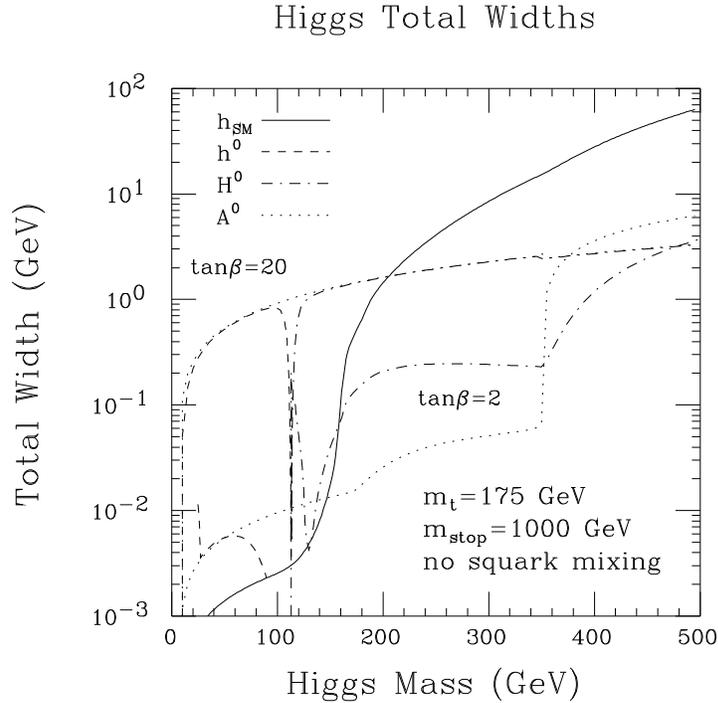}

\medskip
\caption{Total width versus mass of the SM and MSSM Higgs bosons
for $\mt=175\gev$.
In the case of the MSSM, we have plotted results for
$\tan\beta =2$ and 20, taking $\mstop=1\tev$ and
including two-loop radiative corrections, 
neglecting squark mixing; SUSY decay channels are assumed to be  
absent.}
\end{figure}

\begin{figure}[t]
\centering
\hspace{0in}\epsfxsize=4in\epsffile{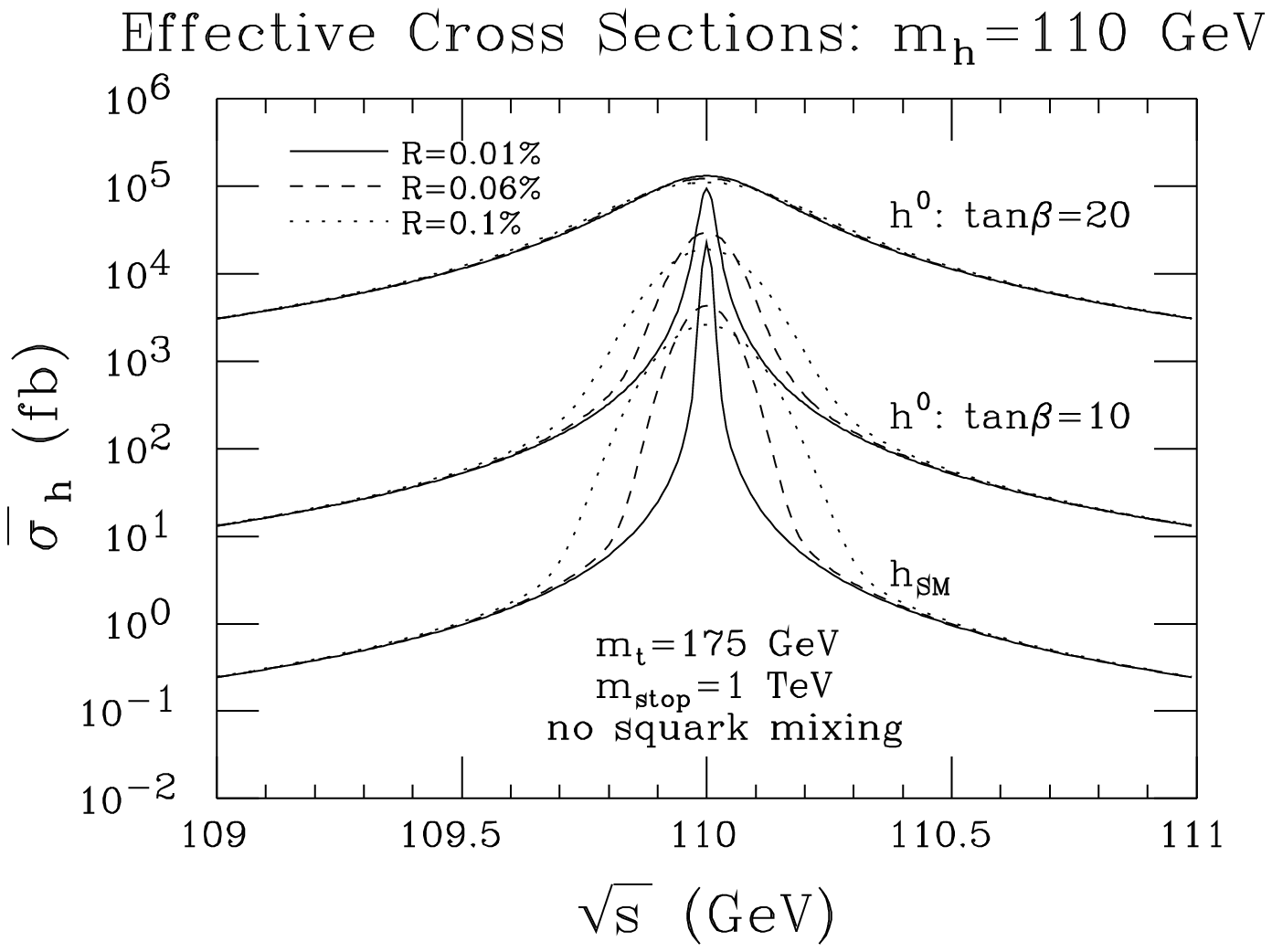}

\medskip
\caption{The effective cross section, $\sighbar$, 
obtained after convoluting $\sigma_{\h}$ with 
the Gaussian distributions for $R=0.01\%$, $R=0.06\%$, and $R=0.1\%$, is
plotted as a function of $\protect\rts$ taking $\mh=110\gev$.
Results are displayed in the cases: $\hsm$,
$\hl$ with $\tanb=10$, and $\hl$ with $\tanb=20$.
In the MSSM $\hl$ cases, two-loop/RGE-improved radiative corrections
have been included for Higgs masses, mixing angles, and self-couplings
assuming $\mstop=1\tev$ and neglecting squark mixing.
The effects of bremsstrahlung are not included in this figure.}
\end{figure}

\subsection{Energy scan}

An energy scan can be made for a $s$-channel Higgs resonance in the mass range 63~GeV${} < m_h < 2M_W$. For detection we require a significance $S/\sqrt B > 5$ and assume excellent resolution $R=0.01\%$ ($\sigma_{\sqrt s} = 7$~MeV) to assure sensitivity to the narrow width of a SM Higgs. The necessary luminosity per scan point for representative $m_h$ values is
\[ \arraycolsep=1.5em\begin{array}{cl}
\underline{L/\rm point} & \multicolumn{1}{c}{\underline{\quad m_h\quad}\quad}\\
0.01\rm~fb^{-1} & \mbox{110 GeV}\\
0.1\rm~fb^{-1} & \mbox{except near }M_Z\\
1\rm~fb^{-1} & \mbox{near  }M_Z
\end{array}\]
The number of scan points needed to zero in on $m_{h_{\rm SM}}$ within 1~rms spread $\sigma_{\sqrt s} = 7$~MeV  is
\[ \arraycolsep=1.5em\begin{array}{ccc}
\underline{\mbox{\# scan points}} & \underline{\quad L_{\rm total}\quad} & \underline{\quad\delta m_h\quad}\\
230 & 2.3\rm\ fb^{-1} & \mbox{800 MeV (LHC)} \\
3 & 0.03\rm\ fb^{-1} & \mbox{20 MeV }(\ell^+\ell^-\to Zh)
\end{array}\]
where the right-hand column is the assumed prior knowledge on $\delta m_h$.

\subsection{Fine Scan}

Once a rough scan has determined the Higgs mass to an accuracy
\begin{equation}
\delta m_h = \sigma_{\sqrt s} \,d \quad {\rm with}\ d \alt 0.3 
\end{equation}
a three-point fine scan can pinpoint the Higgs mass to still higher precision. For this, the optimal distribution of luminosity is $L$ on the peak location found in the rough scan and $2.5L$ on each of the resonance wings at $\pm2\sigma_{\sqrt s}$ from the peak. The measurement of $\sigma_{\rm wings} / \sigma_{\rm peak}$ improves the $m_h$ precision and measures $\Gamma_h$. As an illustration we consider $m_{h_{\rm SM}} = 110$~GeV for which $\Gamma_{h_{\rm SM}} = 3$~MeV. Then with a total luminosity $L_{\rm total} = 3$~fb$^{-1}$ and a resolution $R=0.01\%$, a fine  scan would yield
\begin{equation}
\delta m_h = 0.4{\rm\ MeV} \qquad \delta\Gamma_h = 1\rm\ MeV\,,
\end{equation}
which is a 30\% measurement of the SM Higgs width.

\subsection{$h_{\rm MSSM}$ or $h_{\rm SM}$?}

After the Higgs discovery, the next burning question is whether it is the SM Higgs or the MSSM Higgs boson. There are two ways to know. The first way is to measure $\Gamma_h$ and $\Gamma(h\to \mu\mu)\times B(h\to \bar bb)$. The couplings of the MSSM Higgs to $\bar bb$ and $\mu^+\mu^-$ are substantially greater than the SM Higgs coupling to a heavy MSSM Higgs mass of $m_H \sim 400$~GeV. The two scenarios are thereby distinguishable at the $3\sigma$ level with $L = 50$~fb$^{-1}$ and $R=0.01\%$, except for $m_h$ near $M_Z$.

The second way is to find the other heavier MSSM Higgs bosons. At the LHC there are some regions of $\tan\beta$ vs.\ $m_A$ where only the lightest MSSM Higgs boson can be discovered. In the larger $m_A$ limit of many supergravity models, the masses of $H^0$, $A^0$, and $H^\pm$ are approximately degenerate and $h$ looks increasingly like $h_{\rm SM}$ in its properties. There are 3 possible $H^0, A^0$ search techniques at muon colliders:

\begin{enumerate}

\item
\underline{Scan for $s$-channel Higgs}\\[2mm]
With $L_{\rm total} = 50$~fb$^{-1}$ the $H^0, A^0$ discovery prospects are robust for $250{\rm\ GeV} \leq m_{H^0,A^0} \alt \sqrt s$ and $\tan\beta \agt 3$. Overlapping $H^0,A^0$ resonances can be separated by the scan; see Fig.~4. The $H^0,A^0$ widths ($\Gamma \sim 0.1$ to 0.6~GeV) are larger than resolution and can be measured by the scan. 

\begin{figure}[h]
\centering
\hspace{0in}\epsfxsize=4in\epsffile{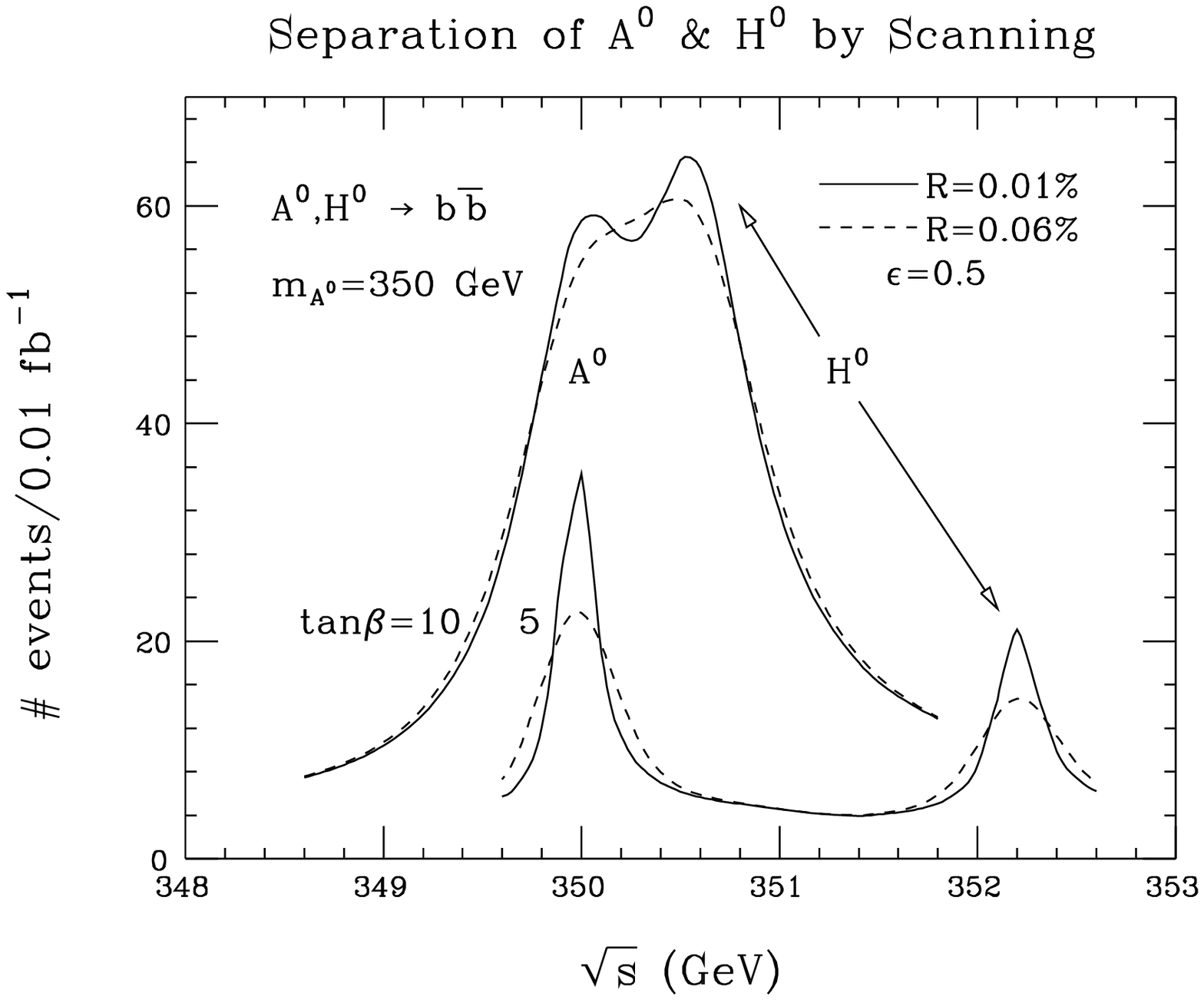}

\medskip
\caption{Plot of $b\anti b$ final state
event rate as a function of $\protect\rts$ for $\mha=350\gev$,
in the cases $\tanb=5$ and 10, resulting from 
the $\hh,\ha$ resonances and the $b\anti b$ continuum background.
We have taken $L=0.01\fbi$ (at any given $\protect\rts$), 
efficiency $\eps=0.5$, $\mt=175\gev$, and
included two-loop/RGE-improved radiative corrections 
to Higgs masses, mixing angles and self-couplings using $\mstop=1\tev$
and neglecting squark mixing. SUSY decays are assumed to be absent.
Curves are given for two resolution choices: $R=0.01\%$ and $R=0.06\%$}
\end{figure}

\item
\underline{Bremsstrahlung tail}\\[2mm]
When the muon collider is run at full energy, $s$-channel production of $H^0, A^0$ will result from the luminosity in the bremsstrahlung tail; see Fig.~5. This production is competitive with the scan search for $\tan\beta\agt5$--7 and invariant mass resolution $\Delta M_{b\bar b} = \pm 5$~GeV.

\begin{figure}[h]
\centering
\hspace{0in}\epsfxsize=4in\epsffile{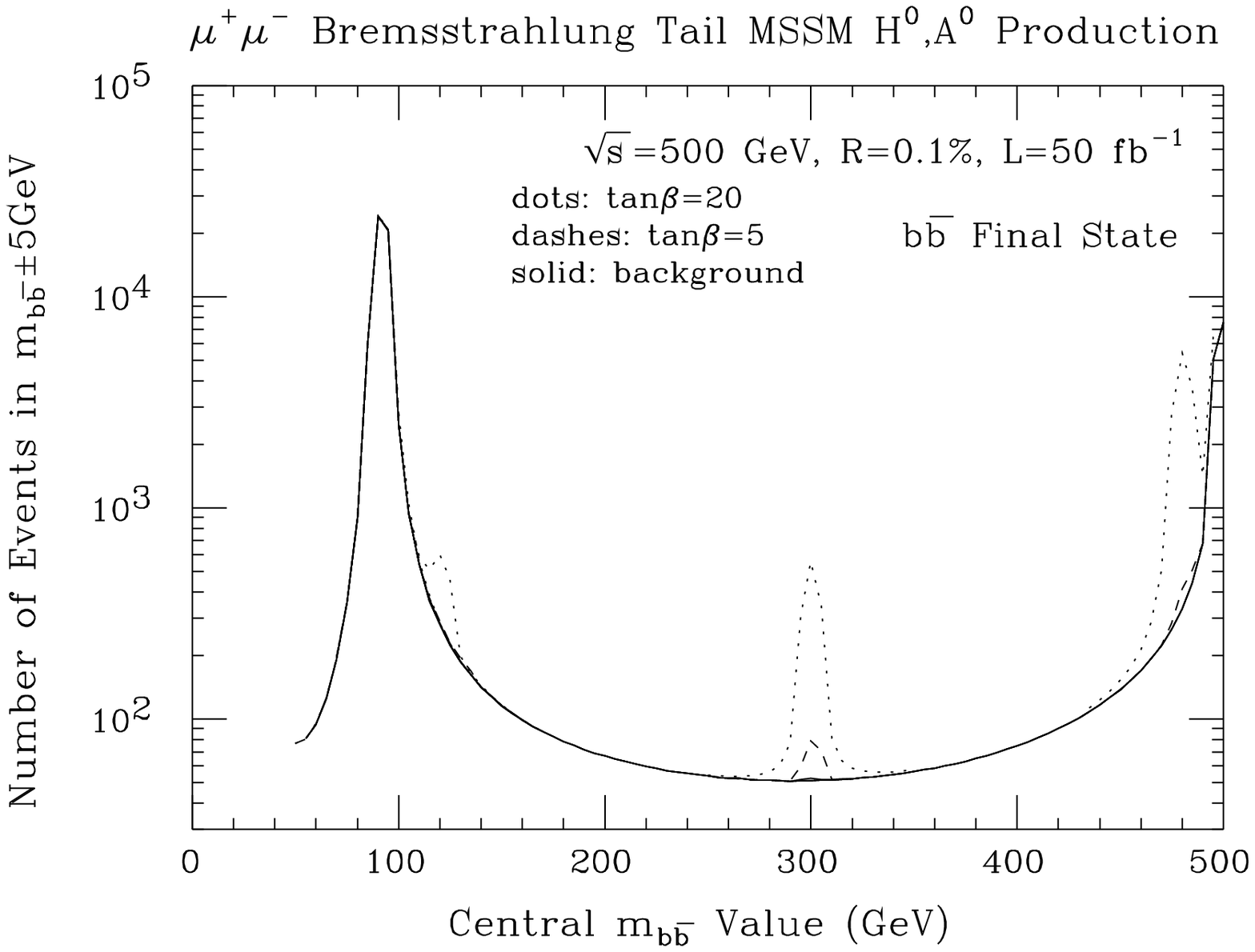}

\medskip
\caption{Taking $\protect\rts=500\gev$, integrated luminosity $L=50\fbi$,
and $R=0.1\%$, we consider the $b\anti b$ final state and
plot the number of events in the interval $[\mbb-5\gev,\mbb+5\gev]$,
as a function of the location of the central $\mbb$ value,
resulting from the low $\protect\rtshat$ bremsstrahlung tail of the  
luminosity distribution.
MSSM Higgs boson $\hh$ and $\ha$ resonances are present for
the parameter choices of $\mha=120$, $300$ and $480\gev$,
with $\tanb=5$ and $20$ in each case. Enhancements for $\mha=120$,
$300$ and $480\gev$ are visible for $\tanb=20$; $\tanb=5$ yields
visible enhancements only for $\mha=300$ and $480\gev$.
Two-loop/RGE-improved radiative corrections are included,
taking $\mt=175\gev$, $\mstop=1\tev$ and neglecting squark mixing.
SUSY decay channels are assumed to be absent.}
\end{figure}

\item
\underline{$HA, H^+H^+$ pair production}\\[2mm]
At the NMC (4~TeV) the discovery a very heavy Higgs boson is feasible via the the processes $\mu^+\mu^-\to Z^*\to HA, H^+H^+$. Cross sections are illustrated in Fig.~6. Once discovery is made, special storage rings can be constructed with c.m.\ energy $\sqrt s \sim m_A, m_H$ to measure the Higgs widths and partial widths.

\bigskip

\begin{figure}[t]
\centering
\hspace{0in}\epsfxsize=4in\epsffile{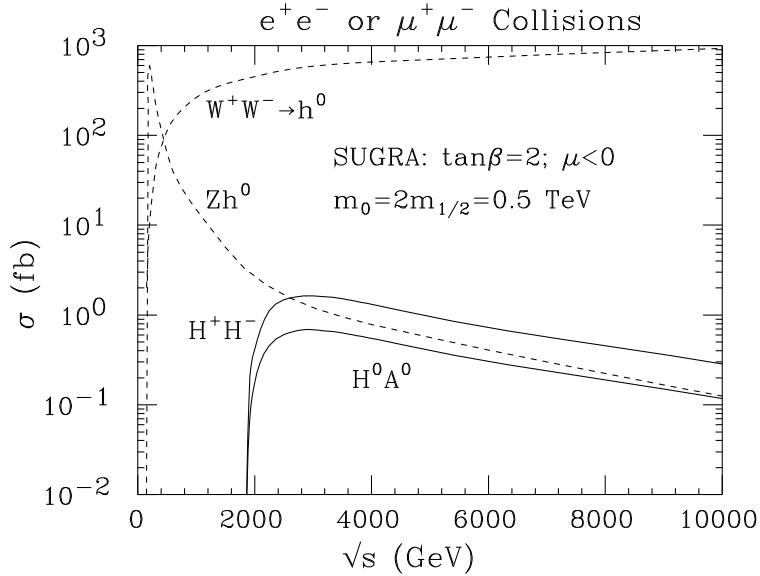}

\medskip
\caption{Pair production of heavy Higgs bosons at a high energy lepton collider. For comparison, cross sections for the lightest Higgs boson production via the Bjorken process $\mu^+\mu^-\to Z^*\to Zh^0$ and via the $WW$ fusion process are also presented.}
\end{figure}

\end{enumerate}

\noindent
Note that at the NMC the large event rates for production of the light Higgs boson may allow measurement of rare decay modes there, e.g.\ $h\to\gamma\gamma$.

\newpage
\section{Advantages/Necessity of a High Energy Muon Collider}

A compelling case for building a 4~TeV NMC exists for both the weakly or strongly interacting electroweak symmetry breaking scenarios.

\subsection{Weakly interacting scenario\protect\cite{sanfran}}

Supersymmetry has many scalar particles (sleptons, squarks, Higgs bosons). Some or possibly many of these scalars may have TeV-scale masses. Since spin-0 pair production is $p$-wave suppressed at lepton colliders, energies well above the thresholds are {\em necessary} for sufficient production rates; see Fig.~7. Moreover, the single production mechanisms at lepton colliders and the excellent initial state energy resolution are {\em advantageous} in reconstructing sparticle mass spectra from their complex cascade decays. 

\begin{figure}[t]
\centering
\hspace{0in}\epsfxsize=3in\epsffile{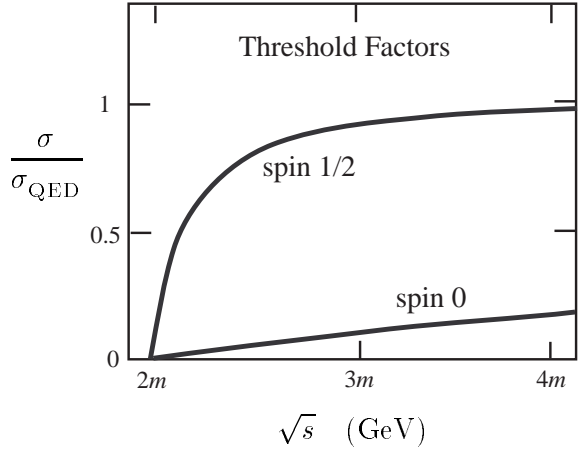}

\medskip
\caption{Comparison of kinematic suppression for fermion pairs and squark pair prodution at $e^+e^-$ or $\mu^+\mu^-$ colliders.}
\end{figure}

\subsection{Strongly interacting electroweak scenarios (SEWS)\protect\cite{sews}}

If no Higgs boson exists with $m_h < 600$~GeV, then partial wave unitarity of $WW\to WW$ scattering requires that the scattering be strong at the 1--2~TeV energy scale. The $WW\to WW$ scattering amplitude is
\begin{eqnarray}
A &\sim& m_H^2 / v^2 \qquad \mbox{if light Higgs} \,,\\
  &\sim& s_{WW} / v^2 \hskip1.5em \mbox{if no light Higgs} \,.
\end{eqnarray}
Then new physics must be manifest at high energies. Energy reach is a critical matter here with subprocess energies $\sqrt{\vphantom1s_{WW}} \agt 1.5$~TeV needed to probe strong $WW$ scattering. Since $E_\mu \sim (\mbox{3--5}) E_W$, this condition implies
\begin{equation}
\sqrt{\vphantom1s_{\mu\mu}} \sim (\mbox{3--5}) \sqrt{\vphantom1s_{WW}} \agt 4\rm\ TeV \,.
\end{equation}
Thus the NMC would have sufficient energy for study of the SEWS.

The nature of the underlying physics will be revealed by the study of all possible vector boson--vector boson scattering channels, since the sizes of the signals depend on the resonant or non-resonant interactions in the different isospin channels; see Table~1.

\begin{table}[h]
\caption{Sizes of SEWS signals in vector boson scattering channels: L (large), M (medium), S (small).}
\medskip
\def\arraystretch{1.4}
\begin{tabular}{c|ccc}
\noalign{\vskip-1.5ex}
final& resonant& resonant& non-resonant\\[-.8ex]
state& scalar ($H^0$)& vector ($\rho_{\rm TC}$)& (LET)\\
\hline
$W^+_L W^-_L$& L& L& S\\
$Z_LZ_L$& L& S& M\\
$W^\pm_L W^\pm_L$& S& M& L\\
$W^\pm_L Z$& S& L& S
\end{tabular}
\end{table}

With 1000~fb$^{-1}$ per year the NMC will allow comprehensive studies to be made of any SEWS signals. First, the vector resonance signals will be spectacular, as illustrated in Fig.~8. The production proceeds via vector meson dominance diagrams. On resonance, $\sqrt s \approx M_V$, the muon collider is a $V$-factory. Similarly, $Z'$ states would also give large signals\cite{sanfran}. Off-resonance production of a vector resonance ($\sqrt s >M_V$) can be detected via the bremsstrahlung luminosity. Second, the scalar particle ($H$) signals will be impressive.
Figure~9 shows the signal
\begin{equation}
\Delta \sigma = \sigma (m_H = 1{\rm\ TeV}) - \sigma(m_H=0)
\end{equation}
expected from a 1~TeV scalar resonance. The signal cross sections are
\begin{equation}
\Delta\sigma(WW) = 70{\rm\ fb\quad and\quad} \Delta\sigma(ZZ) = 40\rm\ fb\,.
\end{equation}
Measurements could differentiate scalar models by measuring the resonance width to $\pm30$~GeV. Finally, there will be good low energy theorem (LET) signals too. 

\begin{figure}[t]
\centering
\hspace{0in}\epsfxsize=4.3in\epsffile{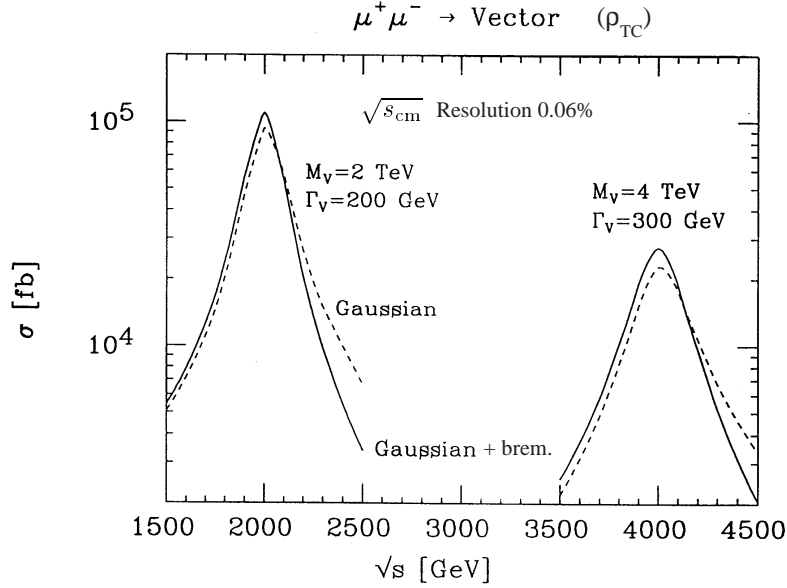}

\medskip
\caption{High event rates are possible if the muon collider energy is set equal to the vector resonance ($Z'$ or $\rho_{\rm TC}$) mass. Two examples are shown here with $R=0.06\%$.}
\end{figure}

Angular distributions of the jets in the $WW\to 4\,$jet final state will provide a powerful discrimination of SEWS from the light Higgs theory, as illustrated in Fig.~10. Here $\theta^*$ is the angles of $q$ and $\bar q$ from $W$-decays in $W$-rest frames, relative to the $W$-boost direction in the $WW$ c.m.\ averaged over all configurations.

The $W_L^+W_L^-\to \bar tt$ channel is another valuable domain for SEWS studies, since  $W_L^+W_L^-\to \bar tt$ also violates unitarity at high energies. Figure~11 illustrates expected cross sections.

\begin{figure}[t]
\centering
\hspace{0in}\epsfxsize=3.5in\epsffile{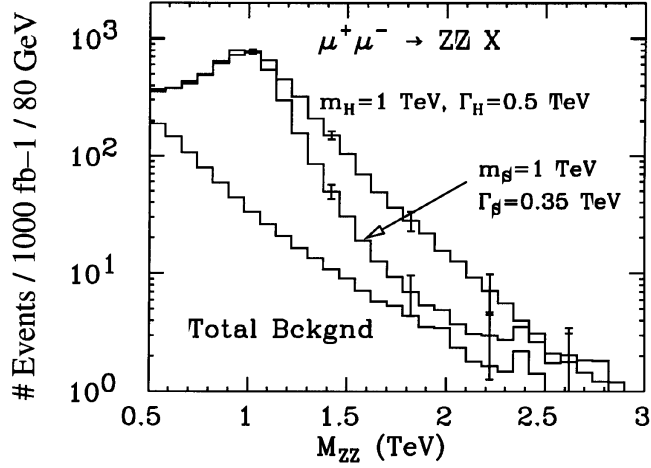}

\medskip
\caption{Events vs.\ $M_{VV}$ for two SEWS models (including the combined backgrounds) and for the combined backgrounds alone in the $ZZ$ final states. Signals shown are: (i) the SM Higgs with $\mhsm=1\tev$, $\Gamma_H=0.5\tev$;
(ii) the Scalar model with $M_S=1\tev$, $\Gamma_S=0.35\tev$.
Results are for $L=1000\fbi$ and $\protect\rts=4\tev$.
Sample error bars  are shown
at $\mVV=1.02$, $1.42$, $1.82$, $2.22$ and $2.62\tev$
for the illustrated 80 GeV bins. Results are for $L=1000$~fb$^{-1}$ and $\protect\sqrt s =4$~TeV.}
\end{figure}

\begin{figure}[h]
\centering
\hspace{0in}\epsfxsize=3.3in\epsffile{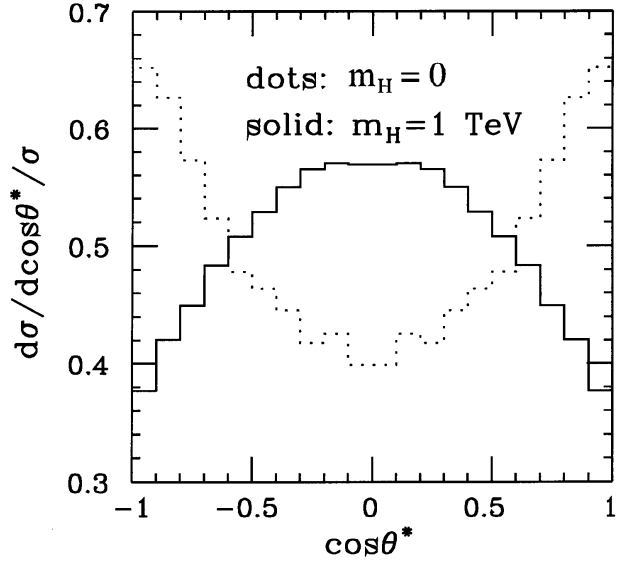}

\medskip
\caption{Plot of normalized cross section shapes and $dN/d\cstar$ (for $L=200\fbi$)
as a function of the $\cstar$ of the $\wp$ decays in
the $\wp\wp$ final state.  Error bars for a typical $dN/d\cstar$
bin are displayed. For this plot we require $\mVV\geq 500\gev$,
$p_T^V\geq 150\gev$, $|\cos\theta_W^{\rm lab}|<0.8$ and
$30\leq p_T^{VV}\leq 300\gev$.}
\end{figure}

\clearpage
\section{Conclusion}

\begin{figure}[t]
\centering
\hspace{0in}\epsfxsize=4in\epsffile{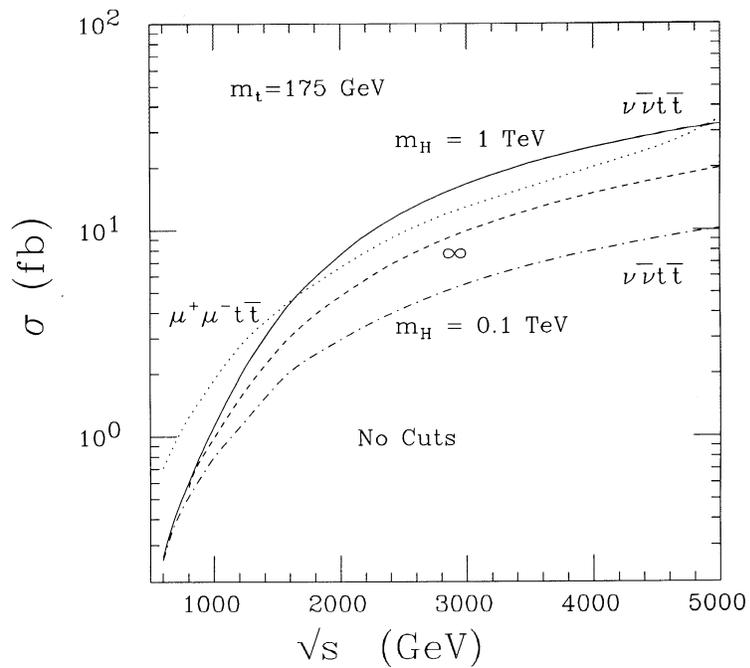}

\medskip
\caption{Cross section vs.\ $\protect\sqrt s$ for $\mu^+\mu^-\to \nu\bar\nu t\bar t, \, \mu^+\mu^- t\bar t$ for Higgs masses $m_H = 0.1$~TeV, 1~TeV, and $\infty$.}
\end{figure}

In summary, muon colliders are a natural match to the physics of electroweak symmetry breaking. The sharp muon beam energy allows

\begin{itemize}

\item precision threshold measurements of $M_W, m_t, m_h$ to test electroweak radiative corrections.

\item $s$-channel resonance scans to precisely determine $m_h$ and measure $\Gamma_h$.

\item discovery and study of the heavy MSSM Higgs bosons in 3 ways.

\end{itemize}

\noindent
The NMC provides the c.m.\ energy and luminosity for

\begin{itemize}

\item heavy supersymmetry thresholds, or

\item SEWS studies.

\end{itemize}

\noindent
If the $WW$ sector proves to be strongly interacting, the NMC is ideally suited to probe the nature of these new interactions.

\break

\section*{Acknowledgements}
This research was supported in part by the U.S.~Department of Energy under Grant No.~DE-FG02-95ER40896 and in part by the University of Wisconsin Research Committee with funds granted by the Wisconsin Alumni Research Foundation.


\begin{references}

\bibitem{private}
R.B.~Palmer, private communication; G.P.~Jackson and D.~Neuffer, private communication.

\bibitem{mwmt} V.~Barger, M.S.~Berger, J.F.~Gunion, and T.~Han, Univ.\ of Wisconsin-Madison report MADPH-96-963 (1996).

\bibitem{zh} V.~Barger, M.S.~Berger, J.F.~Gunion, and T.~Han, Univ.\ of Wisconsin-Madison preprint MADPH-96-979 (1996).

\bibitem{carena} See e.g.\ M.~Carena, J.R.~Espinosa, M.~Quiros, and C.E.M.~Wagner, Phys.\ Lett. {\bf B355}, 209 (1995).

\bibitem{s-channel} V.~Barger, M.S.~Berger, J.F.~Gunion, and T.~Han, Phys.\ Rev.\ Lett.\ {\bf 75}, 1462 (1995);  V.~Barger, M.S.~Berger, J.F.~Gunion, and T.~Han, Univ.\ of Wisconsin-Madison report MADPH-96-930 (1996), to appear in Physics Reports.

\bibitem{parsa} See, for example, Z. Parsa, 
{\it $\mu^+\mu^-$ Collider and Physics
Possibilities}, unpublished.


\bibitem{sanfran}  V.~Barger, M.S.~Berger, J.F.~Gunion, and T.~Han, {\it Proceedings of the Symposium on Physics Potential and Development of $\mu^+\mu^-$ Colliders}, San Francisco, CA, 1995, ed. by D.~Cline and D.~Sanders, Nucl.\ Phys.~B (Proc.\ Suppl.) {\bf 51A}, 13 (1996).

\bibitem{sews} V.~Barger, M.S.~Berger, J.F.~Gunion, and T.~Han, Phys.\ Rev.\ {\bf D55}, 142 (1997).

\end{references}
\end{document}